\documentclass[aps,prl,showpacs,twocolumn,groupedaddress,amssymb]{revtex4}

\usepackage{graphicx}
\usepackage{dcolumn}
\usepackage{bm}

\begin{document}

\title{THz radiation by beating Langmuir waves}

\author{S. Son}
\affiliation{18 Caleb Lane, Princeton, NJ 08540}
\author{Sung Joon Moon}
\affiliation{28 Benjamin Rush Lane, Princeton, NJ 08540}
\author{J.~Y. Park}
\affiliation{EMC2 Fusion Development Corporation}
\date{\today}

\begin{abstract}
An intense terahertz (THz) radiation generated by the beating of two Langmuir waves,
which are excited by the forward Raman scattering, is analyzed theoretically.
The radiation energy per shot can be as high as 0.1 J,
with the duration of 10 pico-second.  
Appropriate plasma density and the laser characteristics are examined.
\end{abstract}

\maketitle

A THz light source has applications in diverse area,
such as the biological diagnostic~\cite{diagnostic},  the planetary
science~\cite{siegel3} and the security scan~\cite{security}.  
In particular, the light source of the frequency range between 1 and 10 THz is
under increasing attention~\cite{siegel, siegel2}. 
There exist technologies to generate the light source of the GHz frequencies 
such as the gyrotron~\cite{gyrotron3,tgyro} and the magnetron~\cite{magnetron},
however, they have the scale problem in the THz range,
unless the magnetic field is ultra-intense~\cite{booske,tgyro}. 
Currently available
technologies, such as the quantum cascade lasers~\cite{qlaser, qlaser3} 
and the free electron lasers~\cite{colson}, suffer from other limitations
in generating an \textit{intense} and \textit{coherent} THz radiation. 
On the other hand, there have been great advances in generating intense lasers
in the visible light and x-ray regimes from the effort in the inertial
confinement fusion~\cite{Fisch, sonlandau, sonbackward,tabak, sonprl, sonpla}.
We note that those technologies could be leveraged to generate an intense,
coherent THz light source, by modulating the frequency to the THz range.

\begin{figure}
\scalebox{0.3}{
\includegraphics{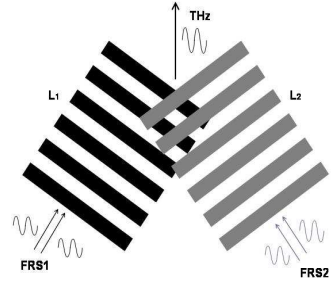}}
\caption{\label{fig1}
A schematic diagram of the THz generation through the interaction of two Langmuir waves
($L_1$ and $L_2$), which are generated in an uniform plasma by the Forward Raman scattering of
a pair of laser pulses (FRS1 and FRS2).
The arrows indicate the wave vector direction.
When the angle between the wave vectors of the Langmuir waves is $\pi/3$,
the waves beat each other to generate a THz radiation (denoted by THz).
See the text for details.
}
\end{figure}

We study one such frequency down-shifting scheme, based on a prudent utilization
of the laser-plasma interaction:
In a plasma of the order of THz plasma frequency, two pairs of intense lasers
excite two Langmuir waves through the Raman scattering~\cite{Rosen}, which subsequently 
lead to an electromagnetic THz wave via the beating current (Fig~{\ref{fig1}}).
In this scheme, the forward Raman scattering (FRS) is preferred to the backward
Raman scattering, as the former converts both the wave vector and the wave frequency down,
while the latter does only the frequency.

Let us start by considering the electromagnetic wave excitation by a pair of Langmuir waves. 
The plasma response is described by the electron continuity and
the momentum balance equations, together with the Maxwell equations:
\begin{eqnarray} 
\frac{\partial n_e }{\partial t} + n_0 \mathbf{\nabla} \cdot \mathbf{v}  &=& 0 \nonumber  \mathrm{,}\\ 
\frac{\partial \mathbf{v} }{ \partial t}  &=& - \frac{e\mathbf{E}}{m_e} \label{eq:momentum} \mathrm{,}\\ \nonumber 
\end{eqnarray}
where $n_e$ and $n_0$ are the perturbed  and the uniform background electron density respectively,
$\mathbf{v}$ is the electron velocity, and $\mathbf{E} $ is the electric field.
The second and the higher order nonlinear terms and
the Lorentz force are ignored in the above equation.
The Maxwell equations are
\begin{eqnarray}
\frac{1}{c}\frac{\partial \mathbf{E}}{ \partial t}  &=& - \frac{4 \pi \mathbf{J}}{c} +  \mathbf{\nabla} \times \mathbf{B} \nonumber \mathrm{,} \\ 
\frac{1}{c}\frac{\partial \mathbf{B}}{ \partial t}  &=& - \mathbf{\nabla} \times \mathbf{E}  
 \label{eq:maxwell} \mathrm{,}\\ \nonumber 
\end{eqnarray}
where $\mathbf{J}$ is the current in the plasma, $c$ is the speed of light,
and $\mathbf{B}$ is the magnetic field.
Eq.~(\ref{eq:maxwell}) can be combined into
\begin{equation}
\frac{1}{c^2}\frac{\partial^2 \mathbf{E}}{ \partial t^2}
+\mathbf{\nabla} \times \mathbf{\nabla} \times \mathbf{E} = -\frac{4\pi}{c^2} \frac{\partial \mathbf{J}}{\partial t} \mathrm{,} \label{eq:maxwell2}
\end{equation}
which leads to the well-known dispersion relation of the electromagnetic wave
in a uniform plasma, $\omega^2 = \omega_{\mathrm{pe}}^2 + (ck)^2$,
where $\omega_{\mathrm{pe}}^2 = 4 \pi n_0 e^2 / m_e$ is the plasma frequency. 
Now, let us consider two Langmuir waves in the plasma.
The perturbed density and the velocity of each Langmuir wave are given as 
\begin{eqnarray}
n_j &=&  L^j n_0 \exp(i \mathbf{k}_j \cdot \mathbf{x} - \omega_j t) \mathrm{,}\nonumber \\
\mathbf{v}_j &=& \frac{\omega_{\mathrm{pe}}^2}{\omega_j}\frac{\mathbf{k}_j}{k_j^2 }    L^j
\exp(i \mathbf{k}_j \cdot \mathbf{x} - \omega_j t) \mathrm{,} \label{eq:langmuir4} \\ \nonumber 
\end{eqnarray}
where the index $j=1,2$ denotes the label for each wave,
and $L^j = \tilde{n}_j/n_0$ is the density perturbation normalized by the background electron density $n_0$.
 In the presence of two Langmuir waves, 
the current $\delta \mathbf{J} $ is given as 
 $\delta \mathbf{J} = - e (n_1 \mathbf{v}_2 + n_2 \mathbf{v}_1) $ 
 with the wave vector 
$\mathbf{k} = \mathbf{k}_1 \pm \mathbf{k}_2$ 
and the frequency $ \omega = \omega_1 \pm \omega_2$.
The Langmuir wave is a longitudinal wave sustained inside the plasma,
which \textit{cannot} excite an E\&M wave by itself. 
However, with two Langmuir waves, 
the wave vector $\mathbf{k}$ is not necessarily parallel with
$\mathbf{k}_1 $ or $\mathbf{k}_2 $, so that the current, $\delta \mathbf{J}$, 
\textit{could} excite an electromagnetic
wave if   the dispersion relation satisfies
\begin{equation}
   |\omega_1 + \omega_2|^2 = c^2|\mathbf{k}_1 -\mathbf{k}_2|^2 + \omega_{\mathrm{pe}}^2 \mathrm{.} \label{eq:dis}
\end{equation}
Note that the frequency of the laser is much higher than that of the Langmuir wave,
and that  $\omega_{1} =  c k_1 $
as $\omega_1 = |\Omega_{a} - \Omega_{b}| \cong c |k_a - k_b| =  c k_1 $,
where $\Omega_a$ and $\Omega_b$ ($k_a$ and $k_b$) are 
the angular frequencies (wave-vectors) of the lasers;
it can be assumed that 
$ \omega_{1} =\omega_2 = ck_1  = ck_2 = \omega_{\mathrm{pe}}$. 
Then Eq.~(\ref{eq:dis}) can be satisfied when the angle between $\mathbf{k}_1 $ and $\mathbf{k}_2$
is $\pi/3$, which can be simplified to
\begin{equation}
\frac{\partial \mathbf{E}}{ \partial t}
+c \frac{\partial}{\partial x} \mathbf{E} \cong   -4 \pi \delta \mathbf{J} \mathrm{.} \label{eq:maxwell5}
\end{equation}
When the angle is not exactly $\pi/3$,  the resonance condition is not exactly satisfied and the  Eq.~(\ref{eq:maxwell5}) should be  modified to 
\begin{equation}
\frac{\partial \mathbf{E}}{ \partial t} + i\delta \omega  \mathbf{E} 
+c \frac{\partial}{\partial x} \mathbf{E} \cong   -4 \pi \delta \mathbf{J} \mathrm{,} \label{eq:maxwell6}
\end{equation}
where $\delta \omega  =  |\omega_1 + \omega_2| - \sqrt{  c^2|\mathbf{k}_1 -\mathbf{k}_2|^2 + \omega_{\mathrm{pe}}^2} $.  If the laser pulse duration $\tau_d$ satisfies $\tau_d \delta \omega < 1 $, the resonance amplification is still relevant.

Let us analyze the excitation of the Langmuir wave by lasers. 
The excitation of each Langmuir wave is described by~\cite{McKinstrie}
\begin{equation}
\left( \frac{\partial }{\partial t} + v_g \frac{\partial}{\partial x} + \nu\right)L^j  = -i2 \omega_{\mathrm{pe}}A_a A^*_b  \mathrm{,} \label{eq:langmuir}
\end{equation}
where the subscripts $a$ and $b$ are the labels for the lasers,
$A_{(a,b)}= eE_{(a,b)}/m_e\Omega_{(a,b)}c$ is 
the electric field amplitude $E_{(a,b)}$
of the two lasers normalized by $ E_C = m_e\Omega_{(a,b)}c/e$, 
$\Omega_j$ is the laser frequency,
$L^j$ is the Langmuir wave amplitude given in Eq.~(\ref{eq:langmuir}),
and $\nu$ is the Landau damping rate which can be ignored here.
The Langmuir wave and the lasers satisfy the conservation relation: 
 $\omega_{\mathrm{pe}} = \Omega_a -  \Omega_b$.
Denoting the laser duration by $\tau_d$, the maximum Langmuir wave intensity is
\begin{equation}
  L^j_{\mathrm{max}} \cong  - 2 \omega_{\mathrm{pe}} \tau_d  |A_aA_b| \mathrm{.} 
\label{eq:lang}
\end{equation}
In order for the resonance amplification of the forward Raman scattering is relevant, the condition $\tau_d \omega_{\mathrm{pe}} > 1 $ should be satisfied, which suggests that the pulse duration $\tau_d$ or the laser interaction time should be longer than the THz radiation oscillation time.  
The total THz radiation can be estimated from Eqs.~(\ref{eq:maxwell5}) and (\ref{eq:lang}).
The maximum beating current, $\delta \mathbf{J} =- e (n_1 \mathbf{v}_2 + n_2 \mathbf{v}_1) $,
is given from Eq.~(\ref{eq:lang}) as 
\begin{equation} 
 J_{\mathrm{max}} \cong en_0c L^1_{\mathrm{max}} L^2_{\mathrm{max}} \mathrm{,} \label{eq:current} \mathrm{.}
\end{equation}
The maximum electric field of the THz radiation, estimated from Eqs.~(\ref{eq:maxwell5}) and (\ref{eq:current}), is
\begin{equation}
   E_{\mathrm{max}} \cong \omega_{\mathrm{pe}} T E_C L^1_{\mathrm{max}} L^2_{\mathrm{max}} \mathrm{.} \label{eq:max}
\end{equation}
where $T = S/c$ is the transit time of the electromagnetic wave across the Langmuir wave-excited
region of the spot size $S$, and  $E_C = m\omega_{\mathrm{pe}}c / e$.
Finally the maximum energy radiated in the THz regime is estimated to be  
\begin{equation}
   E_{\mathrm{total}}  \cong  \frac{ E_{\mathrm{max}}^2 } { 8 \pi }  c S^2 \tau_l, \label{eq:power}
\end{equation}
where  $E_{\mathrm{total}}$  is the total energy
and $\tau_l$ ($> \tau_d$) is the decay time of the Langmuir waves.  

As an example, 
consider  a THz wave in a plasma of $n_0 = 10^{16}  \ \mathrm{cm^{-3}} $,
where $\omega_{\mathrm{pe}} = 5.64 \times 10^{12}  \ \sec^{-1}$, and
two Nd:YAG lasers with the spot size of $S =  0.1 \ \mathrm{cm}$ 
and the wavelength of $1 \ \mu \mathrm{m}$ such that
$c \delta k =\omega_{\mathrm{pe}}$, where $\delta k$ is the wave vector difference of  the laser pair.   
Each set of lasers would excite a Langmuir wave via the FRS.
With the intensity of $ I = 10^{13} \  \mathrm{W} / \mathrm{cm}^2$
and the duration of $ \tau_d = 10^{-11}  \ \sec$,  
the laser would have the energy of 1 J per shot, resulting in the maximum Langmuir wave 
$L_{\mathrm{max}} = 0.5 \times 10^{-3} $ (see Eq.~(\ref{eq:lang})).
The maximum electric field and the total energy is $E_{\mathrm{max}} = 1.2 \times 10^{-5} E_C$
and $E_{\mathrm{total} } = 2.4 \times 10^{-5} \mathrm{J}$ respectively, assuming $\tau_d = \tau_l$
(see Eqs.~(\ref{eq:max}) and (\ref{eq:power})).
 
As one more example, consider the same plasma and the input lasers.
Instead of $I = 10^{13}  \ \mathrm{W} / \mathrm{cm}^2$, let us assume  $I = 10^{14}  \ \mathrm{W} / \mathrm{cm}^2$,
which corresponds to 10 J per shot per laser.
By the same reasoning, we obtain $L_{\mathrm{max}} = 0.5 \times 10^{-2} $,
$E_{\mathrm{max}} = 1.2 \times 10^{-3} E_C$, and  $E_{\mathrm{total}} = 0.24 \ \mathrm{J} $,
assuming $\tau_d = \tau_l$. 
Since the duration is around 10 pico-second,  the temporal radiation power can be larger than 200 GW. 
 The pump-to-terahertz energy  efficiency can be as large as a few percents or even tens of percents for very intense lasers, which is comparable to the gyrotrons in the GHz range~\cite{gyrotron3,tgyro}. It should be noted that the  pump-to-terahertz energy  efficiency is very small for the THz sources.
For example, in the ultra short laser-based source~\cite{tlaser}, the efficiency  is less than 0.1 \%~\cite{tlaser}.  
In the electron beam based source~\cite{beam, beam2}, it is less than 0.01 \%. 
In the quantum cascade lasers~\cite{qlaser, qlaser3}, the efficiency is not even the concern
as the requirement of maintaining the low temperature is a much more severe drawback. 

To summarize, a THz generation scheme based on an articulate use of
four intense laser pulses is considered.
In this scheme, two pairs of lasers with slightly different frequencies excite 
two Langmuir waves in a plasma,
which beat each other to generate a THz radiation.
For this scheme, a uniform dense plasma (of order of $10^{16}\ \mathrm{cm^{-3}}$)
and four intense lasers of a few J per shot are required.
The radiation from this scheme is not readily tunable;
the laser frequency or the plasma density should be adjusted for a specific frequency.
Furthermore, the Langmuir waves needs to be excited close to the wave breaking limit,
which could impose a tough requirement on the plasma uniformity and the temperature. 
However, in comparison with other laser sources, this scheme has the following advantages.
First, the achievable intensity is  high; the radiation energy can be in the range of
1 $\mu\mathrm{J}$ to 1 J per shot in a pico-second duration.
The temporal power could easily exceed GW. 
Second, the operation is relatively easy and inexpensive.
For instance, comparing to the the conventional free electron laser,
strong magnets or relativistic electron beams are not required. 

\end{document}